# Design, fabrication and low-power RF measurement of an X-band dielectric-loaded accelerating structure

Yelong Wei, [1,6*] Hikmet Bursali [1,2], Alexej Grudiev [1], Ben Freemire [3], Chunguang Jing [3,4], Joel Sauza Bedolla [5], Rolf Wegner [1], and Carsten Welsch [6]

[1]*CERN, Geneva CH-1211, Switzerland*

[2]*Sapienza University of Rome, Rome, Italy*

[3]*Euclid Techlabs LLC, Bolingbrook, Illinois 60440, USA*

[4]*High Energy Physics Division, Argonne National Laboratory, Lemont, Illinois 60439, USA*

[5]*Lancaster University, Lancaster, UK*

[6]*University of Liverpool and Cockcroft Institute, UK*

Abstract: Dielectric-loaded accelerating (DLA) structures are being studied as an alternative to conventional disk-loaded copper structures to produce the high accelerating gradient. This paper presents the design, fabrication and low-power RF measurement of an externally-powered X-band DLA structure with a dielectric constant $\varepsilon_r = 16.66$ and a loss tangent $\tan\delta = 3.43 \times 10^{-5}$. A dielectric matching section for coupling the RF power from a circular waveguide to an X-band DLA structure consists of a very compact dielectric disk with a width of 2.035 mm and a tilt angle of 60°, resulting in a broadband coupling at a low RF field which has the potential to survive in the high-power environment. Based on simulation studies, a prototype of the DLA structure was fabricated. Results from bench measurements and their comparison with design values are presented. The detailed analysis on the fabrication error which may cause the discrepancy between the RF measurements and simulations is also discussed.

## I. Introduction

Over the past few decades, there has been a major effort to understand the high gradient limits of conventional disk-loaded metal accelerating structures, as well as to develop alternative structures that may be capable of producing high gradient. One promising concept is the dielectric-loaded accelerating (DLA) structures which utilize dielectrics to slow down the phase velocity of travelling waves in the vacuum channel. A DLA structure comprises a simple geometry where a dielectric tube is surrounded by a conducting cylinder. The simplicity of DLA structures offers great advantages for fabrication of high frequency (>10 GHz) accelerating structures, as compared with conventional metallic accelerating structures which demand extremely tight fabrication tolerances. This is of great importance in the case of linear colliders, where tens of thousands of accelerating structures have to be built. Moreover, the relatively small diameter of DLA structures facilitates the use of quadrupole lenses around the structures. The DLAs are also advantageous in terms of the ease of applying damping schemes for beam-induced deflection modes [1-2], which can cause bunch-to-bunch beam breakup and intrabunch head-tail instabilities [3].

The DLA structures were initially proposed in the 1940s [4-7], and experimentally demonstrated in the 1950s [8-10]. Since that time, disk-loaded metallic structures have prevailed for accelerator research and development because of their high quality factor and high field holding capability. Thanks to remarkable progress in new ceramic materials with high dielectric permittivity ($\varepsilon_r > 20$), low loss ($\tan\delta \leq 10^{-4}$) [11-13], and ultralow-loss ($\tan\delta \leq 10^{-5}$) [14-15], studies on DLA structures are gradually being revived. For example, fused silica, chemical vapor deposition (CVD) diamond, alumina and other ceramics have been proposed as materials for DLA structures [16-18], and experimentally tested with high-power wakefield accelerating structures at Argonne National Laboratory [19-22]. In the last two decades, different kinds of DLA structures with improved performance have been reported, such as a dual-layered dielectric structure [23], a hybrid dielectric and iris-loaded accelerating structure [24], a multilayered dielectric structure [25], a disk-and-ring tapered accelerating structure [26], a dielectric disk accelerating structure [27], and a dielectric assist accelerating structure [28-31]. Since a high accelerating gradient of up to 100 MV/m has been demonstrated at room temperature for an X-band copper structure [32-33], high-gradient X-band technology have received considerable attention from the light-source [34-37] and medical [38-40] communities. Building on these developments, a DLA structure operating at a high frequency (X-band) appears to be very promising for future linear accelerators.

We describe in this paper a detailed design, fabrication and low-power RF measurement of an X-band DLA structure. In order to efficiently couple the RF power from a rectangular waveguide to an X-band DLA structure, two modules are adopted (see Fig. 1): a $TE_{10}$-$TM_{01}$ mode converter with choke geometry and a dielectric matching section. The mode converter is used to convert the $TE_{10}$ mode from a rectangular waveguide to the $TM_{01}$ mode in a circular waveguide. The dielectric matching section provides a good match for the impedance of the $TM_{01}$ mode between the circular waveguide and the DLA structure. A choke flange separates the DLA structure from the mode converter and thus makes the mode converter independent of the dielectric properties. In this case, the mode converter can be reused for similar experiments operating at the same frequency. Section II presents detailed RF design of an X-band DLA structure and dielectric matching sections with a dielectric constant $\varepsilon_r = 16.66$ and a loss tangent $\tan\delta = 3.43 \times 10^{-5}$. Section III shows the fabrication process for obtaining the prototypes of mode converters and DLA structure. Section IV presents the low-power RF measurement for the full-assembly prototypes and its comparison with the simulations. Section V investigates the fabrication errors to cause the discrepancy between the measurement and the simulations. Section VI gives the conclusions.

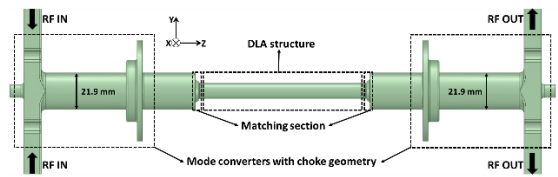

Figure 1. Conceptual illustration of an externally-powered DLA structure connected with two matching sections and two $TE_{10}$-$TM_{01}$ mode converters with choke geometry.

## II. Design of an X-band DLA structure and dielectric matching sections

In this section, the RF properties of an X-band DLA structure (see Fig. 2) are studied. A dielectric matching section to efficiently couple the RF power from a circular waveguide into the DLA structure is also described in detail.

### A. A DLA Structure

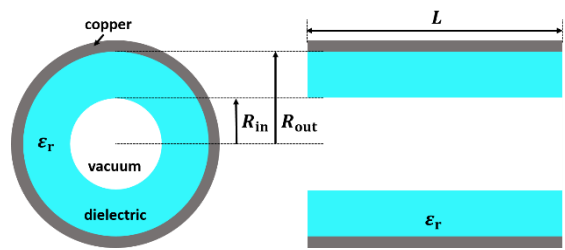

Figure 2. Front view and longitudinal cross section of a cylindrical DLA structure. $\varepsilon_r$, $R_{in}$, $R_{out}$, and $L$ represent dielectric constant, inner radius, outer radius, and length for the DLA structure.

MgTiO$_3$ ceramic, with good thermal conductivity and ultralow power loss, which has been studied in [17], is chosen as the dielectric material for our DLA structure. An accurate measurement of the dielectric properties has to be performed before using such a ceramic for our RF design. As shown in Fig. 3, a $TE_{01\delta}$ silver-plated resonator, which is designed for testing ceramics at an X-band frequency, is used to measure the dielectric constant $\varepsilon_r$ and loss tangent $\tan\delta$ of sample coupons. Four coupons made from the same dielectric rods as for the fabrication are measured. A dielectric constant $\varepsilon_r = 16.66$ and a loss tangent $\tan\delta = 3.43 \times 10^{-5}$ (having error bars 0.6% of the nominal value) are obtained for the RF design of the DLA structure and matching sections which follows.

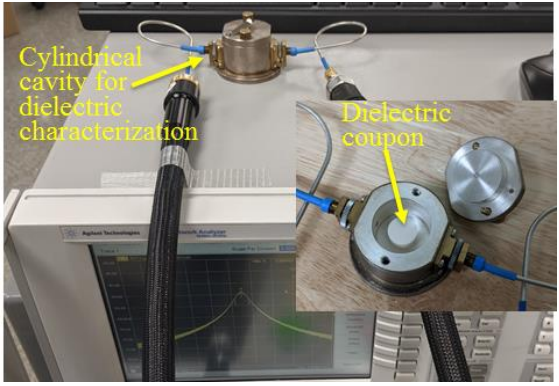

Figure 3. Measurement setup of the dielectric properties of the material sample.

The DLA structure could be potentially used for the CLIC main linac [41-46]. The inner radius is chosen to be $R_{\text{in}} = 3.0$ mm from consideration of the CLIC beam dynamics requirement [43, 44]. The outer radius is then calculated to be $R_{\text{out}} = 4.6388$ mm for an operating frequency of $f_0 = 11.994$ GHz. The group velocity obtained is $v_g = 0.066c$, where $c$ is speed of light. A quality factor of $Q_0 = 2829$ and a shunt impedance of $R_{\text{shunt}} = 26.5$ MΩ m are also derived for such a DLA structure using HFSS [47]. The length of the DLA structure is chosen as $L = 100$ mm for the following simulations and mechanical assembly.

## B. A Chamfered Dielectric Matching Section

Through optimization studies [48], a compact, low-field, broadband matching section with a tilt angle of $\theta = 60°$ is obtained. In realistic fabrication, a sharp dielectric corner easily breaks. In order to prevent such a break, a 45° chamfer with a length of 0.254 mm is added to this corner, as shown in Fig. 4. The shape of the outer metal is also changed by rounding with a fillet radius of $R_r = 0.322$ mm, in order to prevent field enhancement near that area.

Figure 4 (a) shows the calculated electric field distribution for the chamfered dielectric matching section at an input power of 1 W. Figure 4 (b) indicates that the electric fields near that area are much lower than those of the DLA structure. In this case, this dielectric matching section has the potential to withstand high-power test.

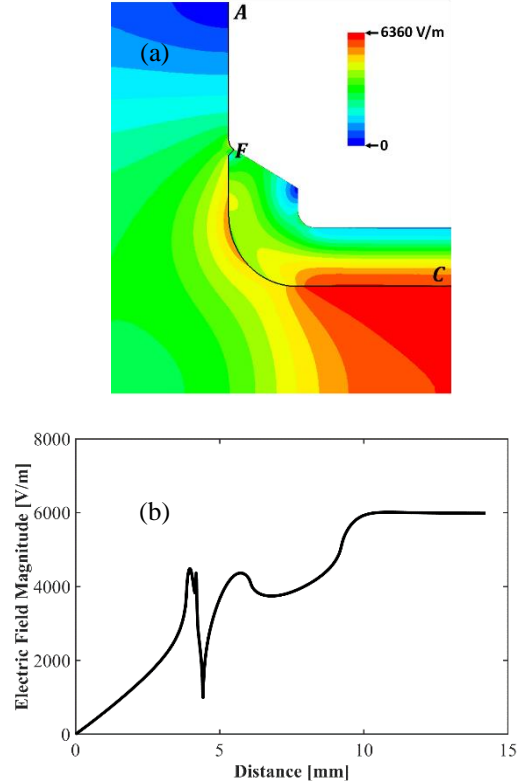

Figure 4. (a) Electric field distribution for the optimum matching section with a chamfered dielectric corner. (b) Electric field magnitude along Line $AFC$ which denotes a section of lines and arcs connected by the points $A$, $F$, and $C$, as shown in (a), where the distance of point $A$ is taken as 0 mm.

Figure 5 shows the calculated $S_{11} = -54$ dB and $S_{21} = -0.03$ dB for the chamfered dielectric matching section at the operating frequency of 11.994 GHz. The $S_{11} = -54$ dB indicates that the reflected RF power is negligibly small. Using $S_{21} = -0.03$ dB, the coupling coefficient is calculated to be 99.3%. This means that almost 100% of RF power is efficiently coupled from the circular waveguide into the DLA structure by using such a dielectric matching section. The $S_{21}$ also has a broad 3 dB bandwidth of more than 1.0 GHz, which allows greater tolerance to potential fabrication errors.

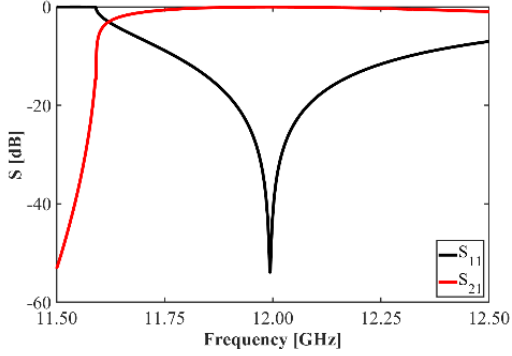

Figure 5: Simulated $S_{11}$ and $S_{21}$ as a function of frequency for the chamfered dielectric matching section.

## C. A Vacuum Microgap

In our realistic fabrication, the entire dielectric tube, including the matching section and the DLA structure, is sintered as a single piece. A thin metallic layer of 0.0508 mm is first coated onto the surface of the whole dielectric tube. The coated dielectric tube is then inserted into the outer copper jacket. However, there is still a microscale vacuum gap caused by metallic clamping between the thin metallic coating and the outer thick copper jacket, as shown in Fig. 6. It is therefore of particular importance to study the dependence of the S-parameters and electric fields on the microgap $d_2$.

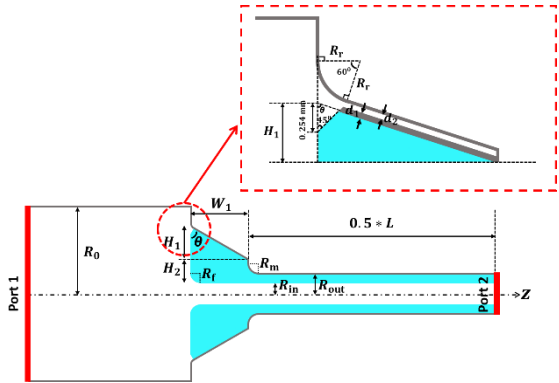

Figure 6. Longitudinal cross section of a circular waveguide, a realistic dielectric matching section with a vacuum microgap, and a DLA structure.

Figure 7 shows how varying the vacuum microgap $d_2$ influences $S_{11}$ and $S_{21}$. With a larger vacuum microgap, $S_{11}$ increases while $S_{21}$ decreases, resulting in worse matching. For a vacuum microgap of $d_2 = 0.2$ mm, $S_{11} = -32$ dB and $S_{21} = -0.32$ dB are obtained, which is still acceptable for our design. However, $S_{11}$ is increased to -28.5 dB and $S_{21}$ remains unchanged, when the vacuum microgap becomes 0.3 mm.

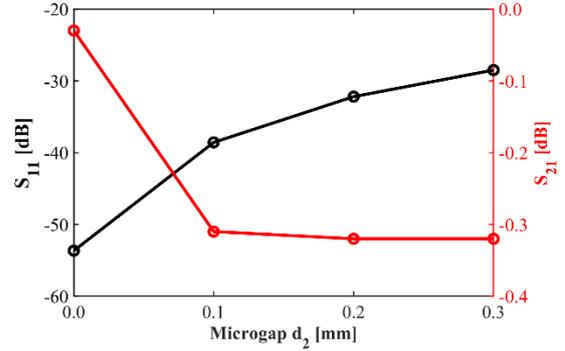

Figure 7. Simulated $S_{11}$ and $S_{21}$ as a function of vacuum microgap $d_2$.

Figure 8 (a) shows the calculated electric field distribution for the realistic matching section with a vacuum microgap, at an input power of 1.0 W. Figure 8 (b) gives the calculated electric field magnitude along Line *AGC* for different vacuum microgaps. There are two peaks in each curve, indicating the relatively strong fields near the chamfered corner and rounding corner, respectively. For a vacuum microgap of 0.3 mm, the peak fields are higher than those of the DLA structure, which may cause arcing in a high-power test. The dielectric matching section is therefore allowed to have a maximum vacuum microgap of 0.2 mm, in which RF fields are still lower than those of DLA structure, $S_{11}$ is better than -30 dB, and the coupling coefficient is 93%. This value is used to guide the fabrication tolerances of the copper jacket and the metallic coating of the dielectric tube.

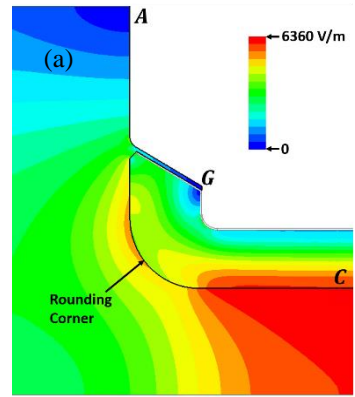

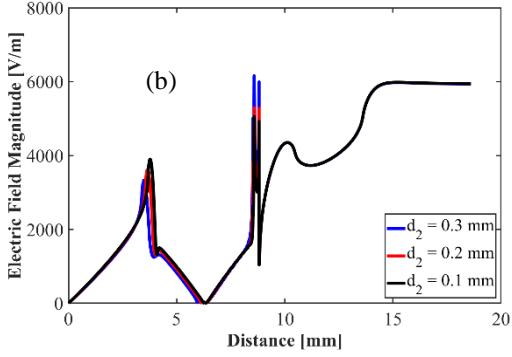

Figure 8. (a) Electric field distribution for the realistic dielectric matching section with a vacuum microgap, where the thin metallic coating is denoted by the white lines. (b) Electric fields magnitude along Line $AGC$ for different vacuum microgaps $d_2$. Here Line $AGC$ denotes a section of lines and arcs connected by the points $A$, $G$, and $C$, as shown in (a), where the distance of point $A$ is taken as 0 mm.

## D. Tolerance Studies

Through previous studies, realistic geometrical parameters for the dielectric matching section and the DLA structure are obtained as follows: $\varepsilon_r = 16.66$, $W_1 = 2.035$ mm, $H_2 = 2.74$ mm, $\theta = 60°$, $R_f = 2.0$ mm, $R_m = 0.5$ mm, $R_{out} = 4.6388$ mm, $R_{in} = 3.0$ mm, and $L = 100$ mm. Using these geometrical parameters, $S_{11} = -54$ dB and $S_{21} = -0.03$ dB are achieved at an operating frequency of 11.994 GHz. The length of the DLA structure does not have any effect on the S-parameters and RF-field performance, so it is ruled out for tolerance studies in this section. The tolerances of key geometrical parameters (see Table 1) are discussed in detail.

Table 1. The tolerances of geometrical parameters for the dielectric matching section and DLA structure.

| $f_0 = 11.994$ GHz | $S_{11} \leq -30$ dB | $S_{11} \leq -25$ dB | $S_{11} \leq -20$ dB |
|---|---|---|---|
| $\varepsilon_r = 16.66$ | [-0.079, +0.081] | [-0.139, +0.148] | [-0.24, +0.27] |
| $W_1 = 2.035$ [mm] | [-0.007, +0.007] | [-0.012, +0.012] | [-0.022, +0.022] |
| $H_2 = 2.74$ [mm] | [-0.015, +0.017] | [-0.027, +0.030] | [-0.051, +0.054] |
| $\theta = 60°$ | [-2.5°, +2.0°] | [-4.3°, +3.7°] | [-7.3°, +7.0°] |
| $R_f = 2.0$ [mm] | [-0.042, +0.040] | [-0.076, +0.068] | [-0.140, +0.120] |
| $R_m = 0.5$ [mm] | [-0.061, +0.049] | [-0.118, +0.090] | [-0.245, +0.151] |
| $R_{out} = 4.6388$ [mm] | [-0.0076, +0.0065] | [-0.0123, +0.0127] | [-0.020, +0.025] |
| $R_{in} = 3.0$ [mm] | [-0.006, +0.007] | [-0.012, +0.012] | [-0.024, +0.020] |

As we know, $S_{21}$ for the realistic matching section has a large 3 dB bandwidth of over 1 GHz, so it is not sensitive to changes in the geometrical parameters. The tolerances are studied by calculating the dependence of $S_{11}$ on the geometrical parameters. By adjusting a certain geometrical parameter from $x$ to $x \pm dx$, $S_{11}$ is calculated and compared with the setting requirements of -30 dB, -25 dB, and -20 dB. As shown in Table 1, $S_{11}$ is very sensitive to $W_1$, $R_{out}$, and $R_{in}$ and less sensitive to $\varepsilon_r$, $H_2$, $\theta$, $R_f$, and $R_m$. The dielectric fabrication accuracy should be better than $\pm 0.02$ mm in order to realize a $S_{11} \leq -20$ dB, which is still acceptable for efficient coupling.

## E. A Full-assembly Structure

In this section, a full-assembly structure (see Fig. 9) is obtained by adding the DLA structure connected together with two matching sections, circular waveguides with the choke geometry, and the $TE_{10}$-$TM_{01}$ mode converters. The mode converters and choke geometry have been studied in Refs. [48-49]. The RF performance of such a full-assembly structure is described in detail. The whole structure is simulated by analysing the electric field distribution and S-parameters from Port 1 to Port 2. RF power loss on both the metallic surface and in dielectrics and accelerating fields in the vacuum channel are also studied in detail.

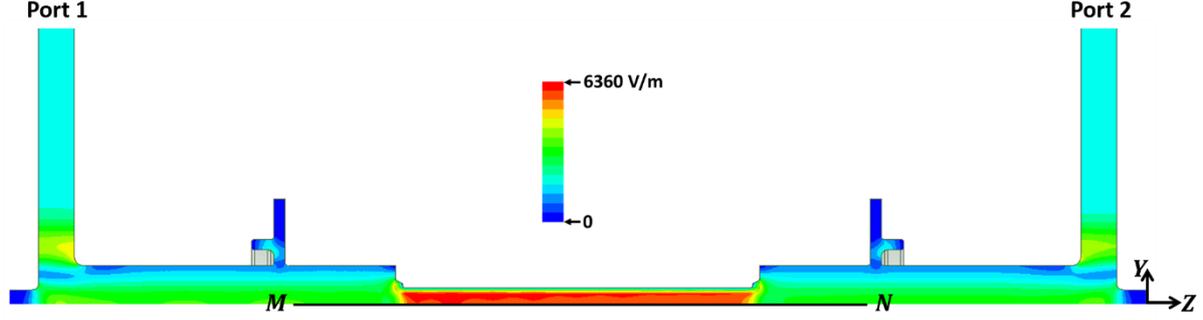

Figure 9. Electric field distribution for the full-assembly structure, where line *MN* is located on the centre along *z*-axis.

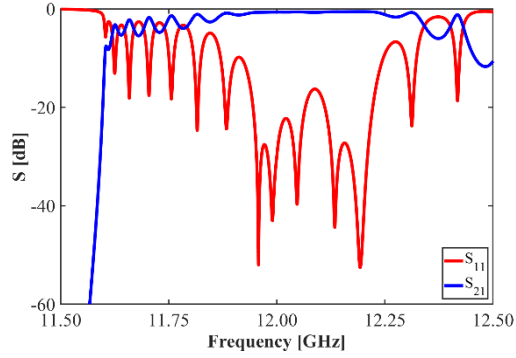

Figure 10. Simulated $S_{11}$ and $S_{21}$ as a function of frequency for the full-assembly structure shown in Fig. 9.

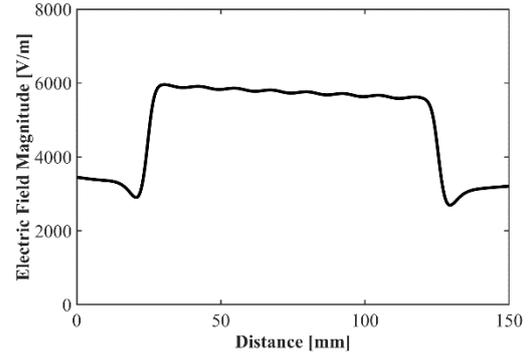

Figure 11. Electric field magnitude along the line *MN* shown in Fig. 9. The distance of point *M* is taken as 0 mm.

Figure 10 gives calculated values of $S_{11} = -40$ dB and $S_{21} = -0.67$ dB for the full-assembly structure at the operating frequency of 11.994 GHz. Using the power density on the metallic surface and in the dielectric area for an input power of 1.0 W, the calculated RF power loss on the metallic surface is $P_{\text{loss\_surface}} = 0.130$ W and the RF power loss obtained in dielectrics is $P_{\text{loss\_dielectric}} = 0.012$ W. So the total RF power loss is $P_{\text{total\_loss}} = 0.142$ W. The output RF power at Port 2 is $P_{\text{out}} = 0.858$ W. We thus achieve a transmission coefficient $S'_{21} = 10 \log(P_{\text{out}}/P_{\text{in}}) = -0.67$ dB, which agrees well with the simulated $S_{21}$ shown in Fig. 22. At the maximum peak power of 40 MW from XBOX [50-51] with a pulse width of 1.5 μs and a repetition rate of 50 Hz an average input power of 3.0 kW will be generated. The power loss on the metallic surface is then 390 W and the power loss in dielectrics is 36 W. A water cooling system is thus required for the high-power test on the full-assembly structure.

Figure 11 shows the electric field magnitude along a line *MN* (see Fig. 9). The electric fields are gradually becoming weaker, due to RF power loss in the dielectric and on metallic surfaces, as the RF fields propagate from point *M* to point *N*. The average accelerating gradient is calculated to be 5773 V/m at an input power of 1.0 W. For a power of 40 MW from XBOX, an average accelerating gradient of 36.5 MV/m can be achieved for our DLA structure.

Figure 12 presents the full-assembly mechanical design for the whole structure. The grey area denotes the outer copper jacket, connected with openings to avoid air trapping when pumping. Two conflat flanges are used to connect the centre part with the end parts, which are $TE_{10}$-$TM_{01}$ mode converters with half-choke geometry.

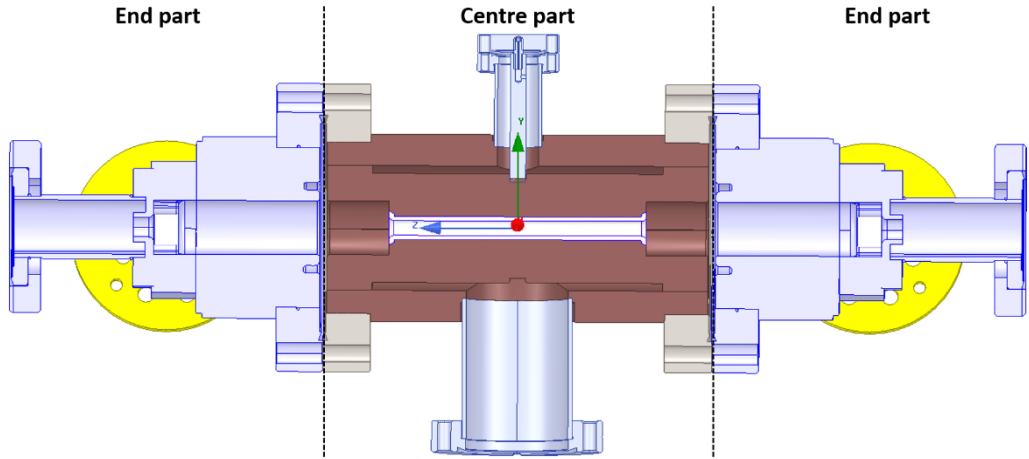

Figure 12. Full-assembly mechanical design for the whole structure, including a centre part and two end parts

## III. Fabrication

### A. TE$_{10}$-TM$_{01}$ mode converter with half-choke geometry (end-part prototype)

The TE$_{10}$-TM$_{01}$ mode converter with half-choke geometry (also called as end-part prototype, see Fig. 12) was fabricated by CERN workshop using conventional welding and brazing. The fabricated end-part prototypes are shown in Fig. 13. Two mode converters with half-choke (see Fig. 13 (a)) which have the exactly the same geometry are used for the full-assembly structure. Fig. 13 (b) shows the copper half-choke geometry and the stainless-steel conflate flange. The screws are used to tight both flanges between centre-part prototype and two end-part prototypes.

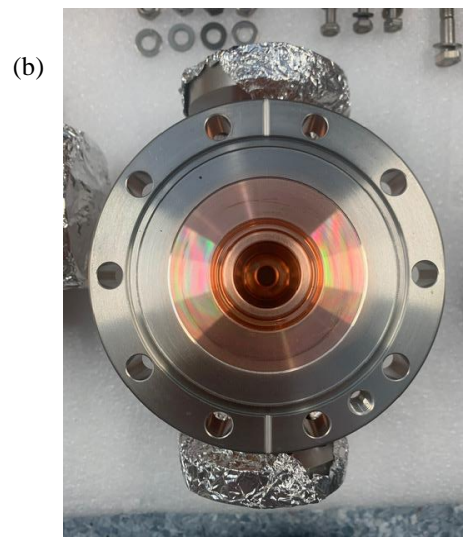

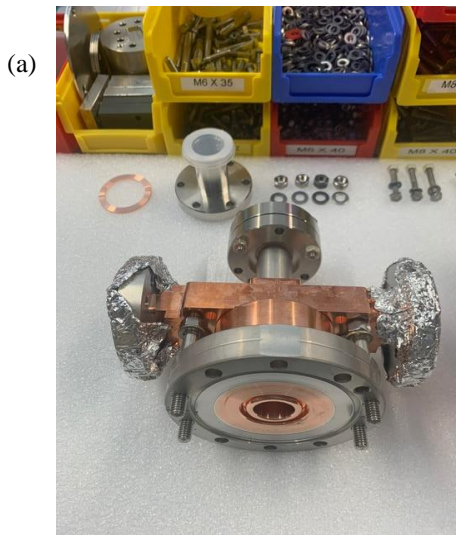

Figure 13. Prototypes of the mode converter with half-choke geometry (a) and conflat flange (b).

### B. Dielectric structure with half-choke geometry (centre-part prototype)

In the realistic fabrication the entire dielectric tube, including the matching section and the DLA structure, is sintered as a single piece. The whole dielectric tube has a thin coating metallic layer of 0.0508 mm. The coated dielectric tube is then inserted into the outer copper jacket, as shown in Fig. 14 (a). The vacuum housing is brazed with the outer copper jacket, forming the centre-part prototype, as shown in Fig. 14 (b). There is a microgap between two-halves copper jacket. Such a microgap may break the symmetry of the whole structure. This centre-part prototype has two conflate flange at both sides which can be tightened together with mode converters to form an ultra-high vacuum.

It should be noted that the outer copper jacket easily slides along the longitudinal direction inside the vacuum housing. In this case, the choke gap $L_c$ (see Fig. 18 in Ref. [48]) at both sides is changed accordingly, which will be discussed in Subsection D of Section IV. Therefore, a stainless-steel stick rod is inserted into the groove of the outer copper jacket in order to stop it from moving along the longitudinal direction. This stick rod is integrated with the flange outside the vacuum housing.

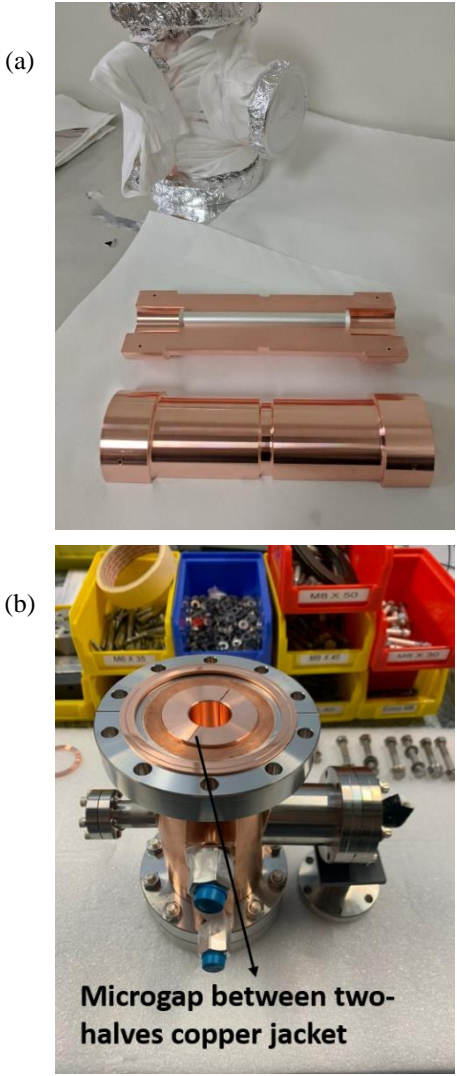

Figure 14. (a) The coated dielectric tube is put into the outer copper jacket; (b) The assembly centre-part prototype with the vacuum housing.

## IV. Low-power RF measurement

In this section, low-power RF measurement is performed for the full-assembly structure by using the Vector Network Analyzer (VNA).

### A. Assembly of two end-part prototypes and an aluminium waveguide

In order to verify the measurement, an aluminium waveguide is fabricated with an inner radius of 10.95 mm. The outer radius of this waveguide is the same as that of the conflate flange. Such an aluminium waveguide is assembled with two end-part prototypes, as shown in Fig. 15 (a). A 4-port VNA is used to measure this assembly structure. Given that the symmetry of the assembly structure, a quarter of the structure is modelled and simulated in HFSS, as shown in Fig. 15 (b).

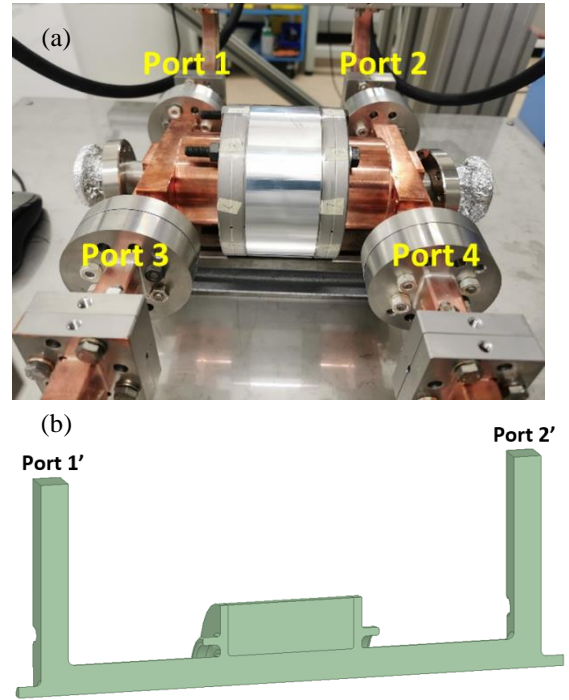

Figure 15. The realistic assembly (a) and simulation modelling (b) of two end-part prototypes and an aluminium waveguide.

The 4-port S-parameters can be transformed into the 2-port S-parameters when the network has a quarter symmetry. Figure 16 shows that the measured $S'_{11} = -34.4$ dB and $S'_{21} = -0.26$ dB while the simulated $S'_{11} = -57.9$ dB and $S'_{21} = -0.05$ dB at the operating frequency of 11.994 GHz. It is found that the measurements are different from the simulations, which may be caused by fabrication errors. However, the measured S-parameters are still acceptable in terms of RF power transmission. Therefore, it is concluded that most of RF power propagates through the end-part prototypes in a normal way.

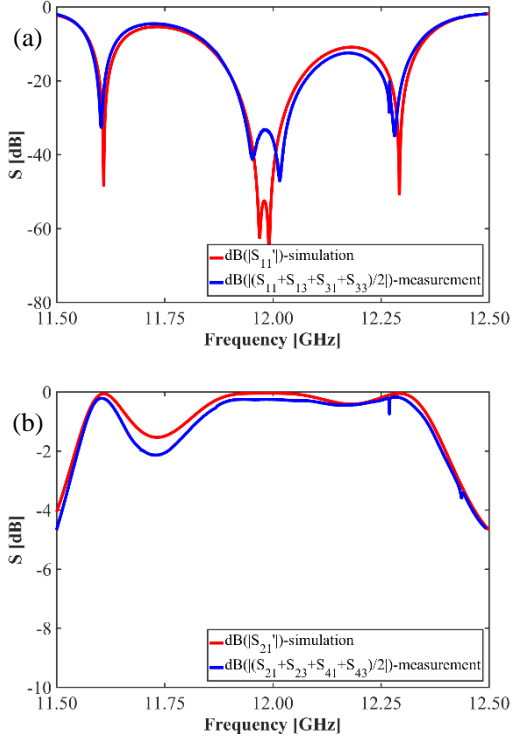

Figure 16. Comparison between simulated and measured S-parameters.

## B. Assembly of two end-part prototypes and centre-part prototype

After confirming the end-part prototypes work well, we move on to assemble two end-part prototypes and centre-part prototypes together, as shown in Fig. 17 (a). Figure 17 (b) shows the modelling of the full-assembly geometry in HFSS simulations, which will be used for comparing with measurements.

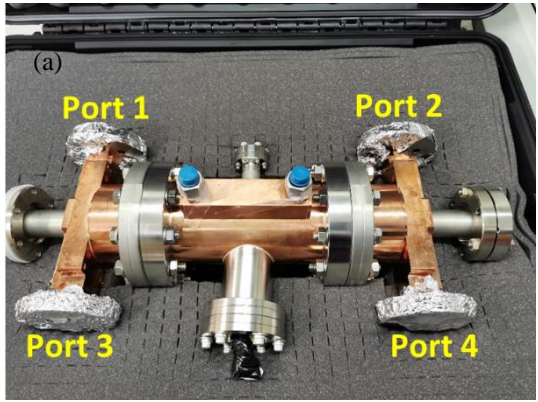

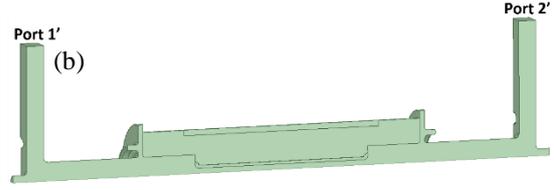

Figure 17. The realistic assembly (a) and simulation modelling (b) of two end-part prototypes and centre-part prototype.

Figure 18 shows that measured $S'_{11} = -11.3$ dB and $S'_{21} = -6.3$ dB while simulated $S'_{11} = -40$ dB and $S'_{21} = -0.67$ dB at the operating frequency of 11.994 GHz. It is obvious that there is a significant discrepancy between measurements and simulations. This discrepancy may be caused by fabrication errors for the DLA structure with the matching sections and microgap between two-halves copper jacket. Investigations into possible reasons for the discrepancy will be discussed in Section V.

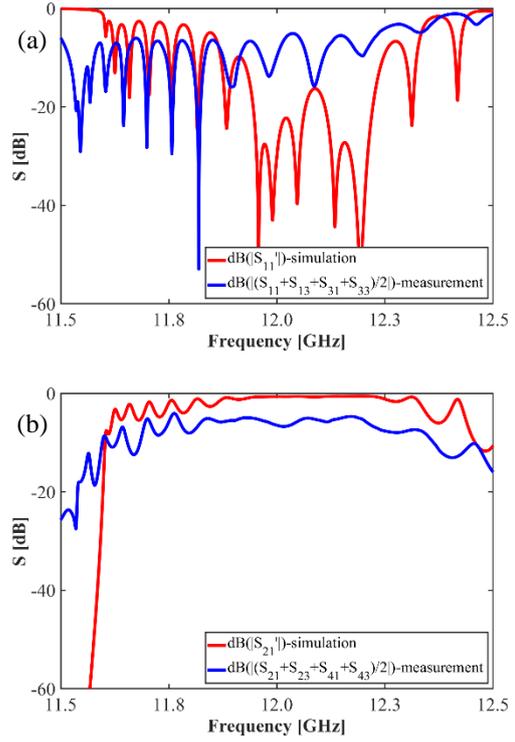

Figure 18. Comparison between simulated and measured S-parameters.

In order to know the electric field distributions along the axis in the full-assembly structure, the bead-pull measurements are carried out as follows. In the measurement, a small dielectric bead attached to a string is pulled on-axis through the whole structure while a low RF power is fed into the structure at the frequency of 11.994 GHz. Dry nitrogen is fed into the structure to avoid permittivity errors due to humidity changes. Electric field profile can be obtained by calculating the change of reflection coefficient with respect to the bead position.

Based on Refs. [52-53], the square of electric field strength $E^2$ is proportional to the change of reflection coefficient $\Delta S = S - S^0$, where $S$ is the reflection coefficient and $S^0$ is the zero-line fitting function. Our full-assembly structure is a 4-port network. The reflection coefficient $S = (S_{11} + S_{13} + S_{31} + S_{33})/2$ for Port 1 and Port 3 while $S = (S_{22} + S_{24} + S_{42} + S_{44})/2$ for Port 2 and Port 4, where $S_{11}$, $S_{13}$, $S_{31}$, $S_{33}$, $S_{22}$, $S_{24}$, $S_{42}$, $S_{44}$ are scattering parameters for our 4-port network. Figure 19 shows the measured electric field distribution on the beam axis of structure at the operating frequency of 11.994 GHz. The red curve denotes the electric field distribution for the case of injecting RF power from Port 1 and Port 3 while the blue curve denotes the electric field distribution for the case of injecting RF power from Port 2 and Port 4. The difference between them demonstrates that the whole structure is asymmetrical because of fabrication error and microgap between two-halves copper jacket. It can also be clearly seen that there are strong reflections at the position of matching sections for both cases. These reflections may also be caused by fabrication errors for the matching sections, which will be discussed in Section V.

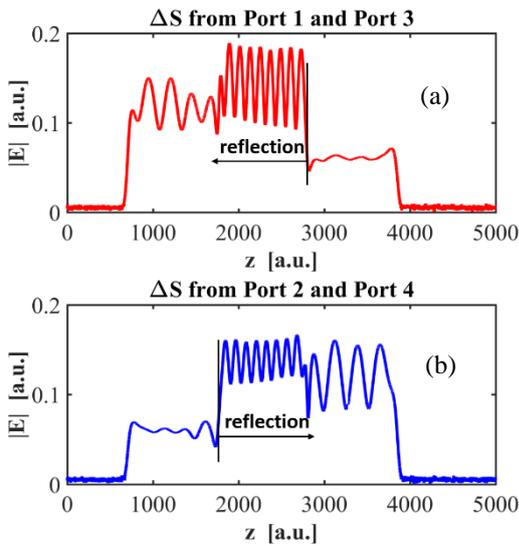

Figure 19. Measured electric field magnitude for injecting RF power from Port 1 and Port 3 (a), and from Port 2 and Port 4 (b), respectively.

## C. Add power splitters for the assembly of two end-part prototypes and centre-part prototype

In previous subsections, we know the fabrication errors and microgap between two-halves copper jacket may break the symmetry of the whole structure. In this case, dipole $TE_{11}$-type modes are excited and propagate through the DLA structure in our previous measurements. This may result in a high power loss. In order to prevent the potential dipole modes and remove the influence of microgap, two power splitters are added into the RF measurement.

At first, two power splitters are added for the assembly of two end-part prototypes and an aluminium waveguide. The 4-port network is changed to a 2-port network. The low-power RF measurement is carried out for this 2-port network, as shown in Fig. 20.

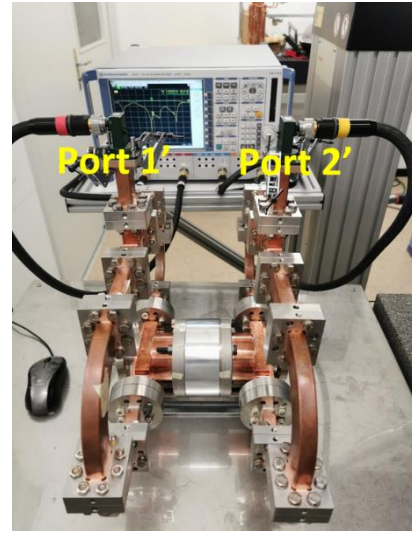

Figure 20. The assembly of two end-part prototype and an aluminium waveguide connected with two power splitters.

Figure 21 shows measured $S'_{11} = S'_{22} = -26.1$ dB and $S'_{12} = S'_{21} = -0.2$ dB at the operating frequency of 11.994 GHz. It can be clearly seen that both $S'_{11}$ and $S'_{12}$ agree well with $S'_{22}$ and $S'_{21}$. This indicates that the power splitters are symmetrical for RF propagating, thereby resulting in a small reflection coefficient and negligible transmission loss. In this measurement, dipole modes are fully suppressed in the structure.

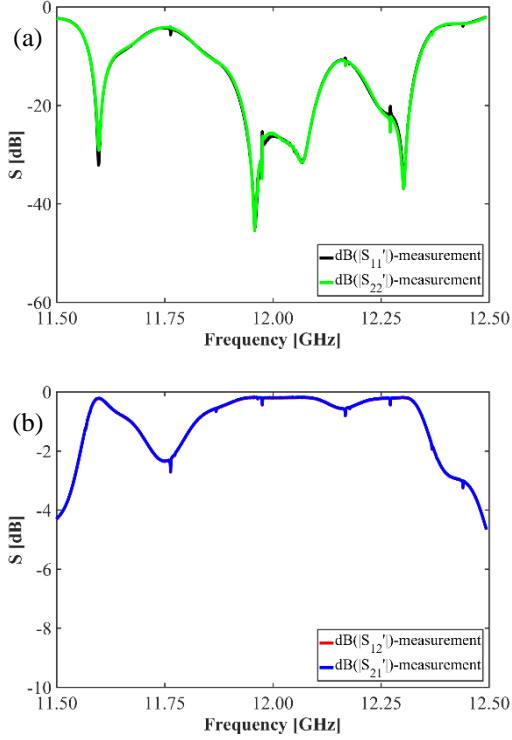

Figure 21. Comparisons between measured (a) $S'_{11}$ and $S'_{22}$, (b) $S'_{12}$ and $S'_{21}$.

After confirming that power splitters work smoothly, we replace the aluminium waveguide with the centre-part prototype which has a DLA structure inside. As shown in Fig. 22, similar RF measurements are carried out for this full-assembly structure.

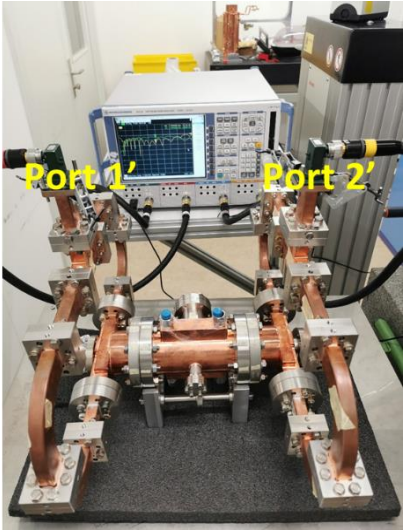

Figure 22. The full-assembly structure connected with two power splitters.

Figure 23 shows measured $S'_{11} = -15.2$ dB, $S'_{22} = -10.2$ dB, $S'_{12} = S'_{21} = -4.96$ dB at the operating frequency of 11.994 GHz. $S'_{11}$ doesn't agree well with $S'_{22}$, which is probably due to asymmetry of the full-assembly structure. Both $S'_{11}$ and $S'_{21}$ are improved as compared to previous 4-port measurements (see Fig. 18). This indicates that symmetrical power splitters are beneficial for suppressing dipole modes propagating in the DLA structure, thereby mitigating the influence caused by microgap between two-halves copper jacket.

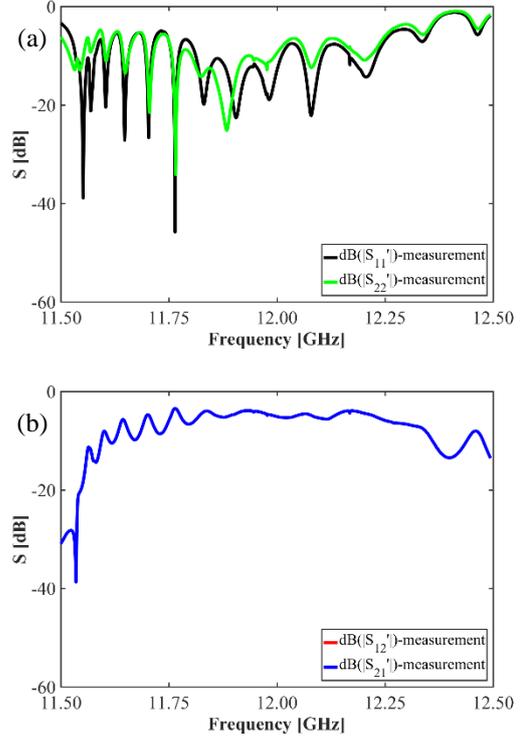

Figure 23. Comparisons between measured (a) $S'_{11}$ and $S'_{22}$, (b) $S'_{12}$ and $S'_{21}$.

## D. Change the choke gap to meet the requirement for high-power test

After adding power splitters for the full-assembly structure, $S'_{11} = -15.2$ dB and $S'_{21} = -4.96$ dB are achieved. Given that this full-assembly structure will be put into XBOX system [50-51] for high-power test in the near future, the reflection coefficient has to be improved better than $S'_{11} \leq -20$ dB in order to protect our XBOX system. In order to meet this requirement, the choke gap at both sides is changed.

At first, we remove the stainless-steel stick rod which should be inserted into the groove of the outer copper jacket in order to stop it from moving. In this situation, the copper jacket enclosing the DLA structure easily slide along the beam axis. If it moves a distance of *d*, the choke gap at one side increases by *d* while the choke gap at the other side decreases by *d* as compared to the original gap of 3 mm, as shown in Fig. 24. It can be seen in Fig. 25 that the measured $S'_{11} = -21.4$ dB when $d = 1.0$ mm although measured $S'_{21} = -7.75$ dB is becoming worse. This means that our full-assembly structure meets the requirement of our XBOX system for high-power test by adjusting the choke gap at both sides.

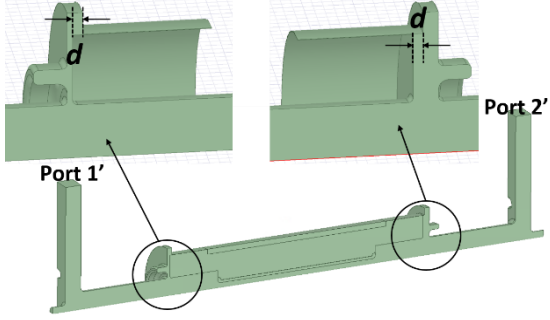

Figure 24. The choke gap at both sides changes with the moving of two-halves copper jacket.

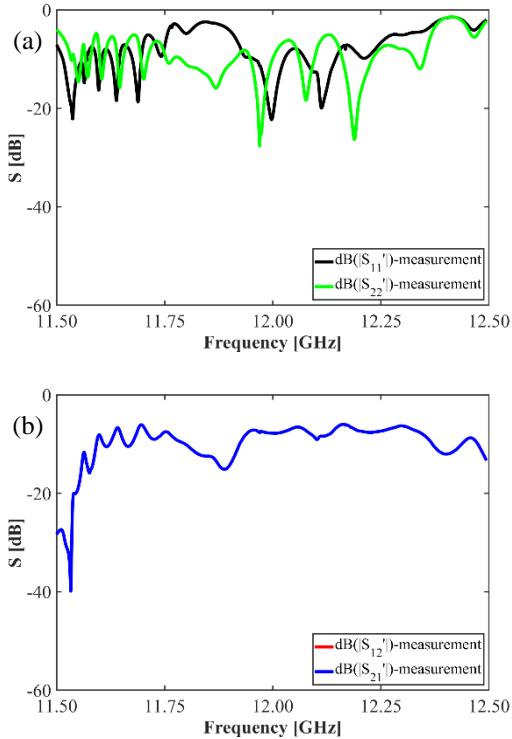

Figure 25. Comparisons between measured (a) $S'_{11}$ and $S'_{22}$, (b) $S'_{12}$ and $S'_{21}$.

## V. Fabrication error analysis

In this section, fabrication error analysis is performed by adjusting the geometry of matching section and the conductivity of coating layer to produce an electric field distribution and S-parameters which are quite similar to those of RF measurements.

It is found that the electric field magnitude and the number of periodic peaks in DLA structure are determined by width $W_1$ of matching section and inner radius $R_{\text{in}}$ of DLA structure in HFSS simulations. When $R_{\text{in}} = 2.99$ mm, the simulated electric field has 7 periodic peaks at DLA structure, which agrees with both curves in Fig. 19. When $W_1 = 2.12$ mm, the simulated electric field magnitude is similar to that of bead-pull measurement from Port 1 and Port 3, as shown in Fig. 26 (a). When $W_1 = 2.08$ mm, the simulated electric field magnitude is also similar to that of bead-pull measurement from Port 2 and Port 4, as shown in Fig. 26 (b).

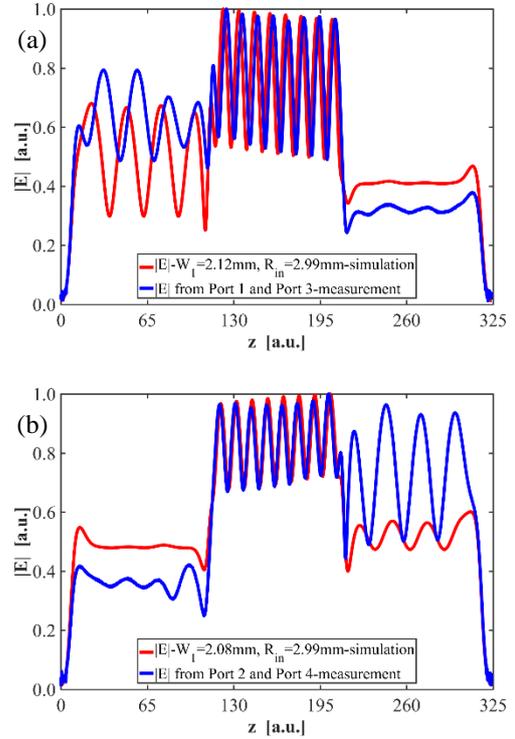

Figure 26. Comparison between simulated and measured electric field magnitude, $W_1 = 2.12$ mm, $R_{\text{in}} = 2.99$ mm (a), $W_1 = 2.08$ mm, $R_{\text{in}} = 2.99$ mm (b).

Therefore, the full-structure modelling (see Fig. 17 (b)) is modified to have a DLA structure with an inner radius of $R_{in} = 2.99$ mm, one matching section with a geometry of $W_1 = 2.12$ mm at one side and the other matching section with a geometry of $W_1 = 2.08$ mm at the other side. The simulated S-parameters are then obtained and compared with the measured S-parameters. Figure 27 shows that simulated $S'_{11} = -11.8$ dB is in good agreement with measured $S'_{11} = -11.3$ dB but simulated $S'_{21} = -0.97$ dB is not consistent with measured $S'_{21} = -6.3$ dB at the operating frequency of 11.994 GHz. This means that the fabrication error on inner radius is 0.01 mm which meets the requirement of fabrication accuracy in subsection of tolerance studies. However, the fabrication error on $W_1$ for one matching section is 0.085 mm and it is 0.045 mm for the other matching section.

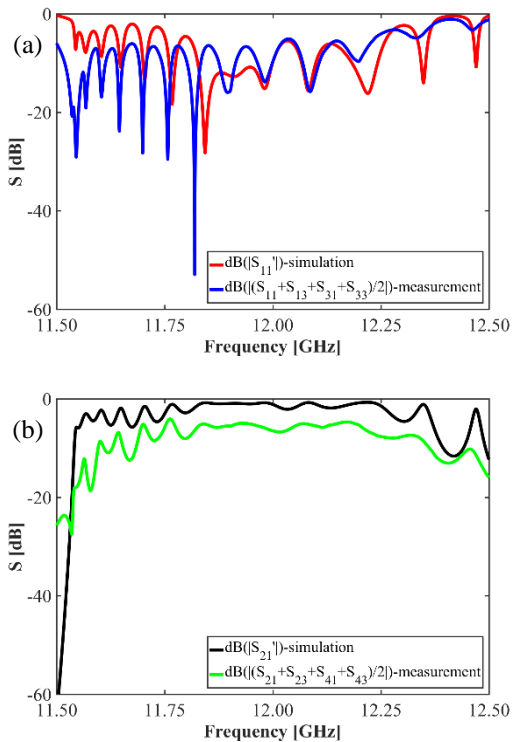

Figure 27. Comparison between simulated and measured S-parameters.

We continue investigating other possible reasons to cause the inconsistence between the simulated and measured $S'_{21}$. Besides the fabrication errors on both $R_{in}$ and $W_1$, when the conductivity of thin coating metal is changed to $\sigma = 5.5 \times 10^5$, the simulated $S'_{21} = -6.25$ dB is almost the same as the measured $S'_{21} = -6.3$ dB at the operating frequency of 11.994 GHz, as shown in Fig. 28. It is also found that the simulated $S'_{11} = -11$ dB has a good agreement with the measured $S'_{11} = -11.3$ dB at the same frequency. This indicates that the whole dielectric tube may have a poor quality of the coating.

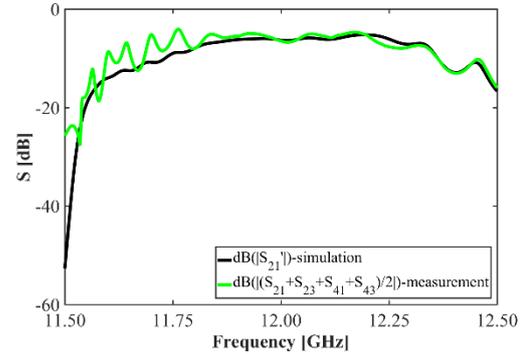

Figure 28. Comparison between simulated and measured S-parameters.

## VI. Conclusions

This paper presented the design, fabrication and low-power RF measurement of an X-band DLA structure at CERN. In order to efficiently couple the RF power from a rectangular waveguide to an X-band DLA structure, the mode converters with choke geometry and the matching sections were carefully designed. Tolerance studies were also carried out for the whole dielectric structure. A full-assembly structure, including the DLA structure connected with two matching sections, $TE_{10}$-$TM_{01}$ mode converters with choke geometry, was analysed in detail.

A prototype of the DLA structure with the matching sections was subsequently built and mechanically assembled with the mode converters for RF measurement. The mode converters were demonstrated to propagate RF power in a normal way. However, a significant discrepancy, due to fabrication error, was found between measured and simulated S-parameters. Geometrical analysis was thus performed to understand the origin of these differences. It was found that the fabrication error on width $W_1$ of the matching sections was the main cause to dilute the reflection coefficient. The poor quality of surface coating on the dielectric tube resulted in a large attenuation for the transmission of RF power. Through adjusting the choke gap, the overall measured reflection coefficient $S'_{11} = -21.4$ dB, which meets the requirement of our XBOX system.

In the next step, the whole structure will be put into Xbox system for high power test. It is foreseen that dielectric RF breakdown [20] and surface resonant multipacting [19-20, 22, 54-55] would be the primary issues to limit the achievable gradient in DLA structures. Thus, further studies are also required to solve these issues for the DLA structures including using an applied axial magnetic field for the whole structure. This will be reported in separate publications.

## Acknowledgments

The authors would like to thank Dr. Walter Wuensch for the useful comments, Dr. Nuria Catalan Lasheras, Serge Lebet, and Sergio Gonzalez Anton for the mechanical and measurement support, the team of Argonne Wakefield Accelerator facility (Dr. Manoel Conde, Dr. John Power, and Dr. Jiahang Shao, etc) for the fruitful discussions, and Dr. Mark Ibison for his careful reading of the manuscript.